\documentclass[aps,prl,twocolumn,superscriptaddress]{revtex4-1}

\usepackage{graphics}
\usepackage{graphicx}
\usepackage{color}
\usepackage{epstopdf}
\usepackage{amsmath}

\usepackage{cancel} 

\begin{document}

\title{Liquid dewetting under a thin elastic film} 

\author{Rafael D. Schulman}
\affiliation{Department of Physics and Astronomy, McMaster University, 1280 Main St. W, Hamilton, ON, L8S 4M1, Canada.}
\author{John F. Niven}
\affiliation{Department of Physics and Astronomy, McMaster University, 1280 Main St. W, Hamilton, ON, L8S 4M1, Canada.}
\author{Michiel A. Hack}
\affiliation{Department of Physics and Astronomy, McMaster University, 1280 Main St. W, Hamilton, ON, L8S 4M1, Canada.}
\author{Christian DiMaria}
\affiliation{Department of Physics and Astronomy, McMaster University, 1280 Main St. W, Hamilton, ON, L8S 4M1, Canada.}
\author{Kari Dalnoki-Veress}
\email{dalnoki@mcmaster.ca}
\affiliation{Department of Physics and Astronomy, McMaster University, 1280 Main St. W, Hamilton, ON, L8S 4M1, Canada.}
\affiliation{Laboratoire de Physico-Chimie Th\'eorique, UMR CNRS Gulliver 7083, ESPCI Paris, PSL Research University, 75005 Paris, France.}

\date{\today}

\begin{abstract}
We study the dewetting of liquid films capped by a thin elastomeric layer. When the tension in the elastomer is isotropic, circular holes grow at a rate which decreases with increasing tension. The morphology of holes and rim stability can be controlled by changing the boundary conditions and tension in the capping film. When the capping film is prepared with a biaxial tension, holes form with a non-circular shape elongated along the high tension axis. With suitable choice of elastic boundary conditions, samples can even be designed such that square holes appear. 
\end{abstract}


\maketitle

Dewetting, the study of the spontaneous withdrawal of a liquid off a substrate, has been a topic of intense research over the last several decades~\cite{Gennes2008,Redon1991,Reiter1992,Reiter2005,Damman2007,Dalnoki1999,Herminghaus2002,Gabriele2006,Redon1994,Fetzer2005,Baumchen2010,McGraw2016,Seemann2001,Seemann2001b,DeSilva2007,Rockford1999,Higgins2000,Konnur2000,Kargupta2002,Suh2002,Sehgal2002}. In part, this is due to the fact that dewetting serves as a powerful tool to probe physical properties and principles such as:   residual stresses in polymer films~\cite{Reiter2005,Damman2007}, rheological properties of viscoelastic materials~\cite{Dalnoki1999,Herminghaus2002,Gabriele2006},  hydrodynamic slip conditions and the dynamics of the contact line~\cite{Redon1994,Fetzer2005,Baumchen2010,McGraw2016}, as well as determining the effective interface potential and Hamaker constants of a system~\cite{Seemann2001,Seemann2001b,DeSilva2007}. In addition, dewetting may be utilized to generate pattern formation at microscopic length scales. Novel dewetting morphologies are typically introduced by chemical and topological patterning in the substrate itself~\cite{Rockford1999,Higgins2000,Konnur2000,Kargupta2002,Suh2002,Sehgal2002}.

Liquid patterning at small length scales may also be achieved using principles of \emph{elastocapillarity}, which is the study of the interplay between a solid's elasticity and a liquid's capillarity. In particular, when a solid substrate is sufficiently deformable (either because it is a soft material or because the chosen geometry is highly compliant), the solid will experience large-scale deformation due to capillary forces of a droplet acting upon it~\cite{Shanahan1987,Marchand2012a,Style2013,Style2013a,Nadermann2013,Park2014,Bostwick2014,Schulman2016,Schulman2017}. Due to this principle, droplets can migrate towards regions of a substrate that are less stiff, which can be used to pattern a soft substrate with liquid droplets~\cite{Style2013}.  In another study, droplets are shown to map out the stresses in free-standing elastic films by assuming a shape which is elongated along the direction of highest tension~\cite{Schulman2017}. More fundamentally, the replacement of a rigid boundary condition with a compliant one leads to unique wetting properties which show departures from the classic descriptions, such as Young's law of partial wetting~\cite{Marchand2012a,Style2013a,Park2014,Schulman2016,Schulman2017}.

Few studies have investigated the intersection of dewetting and elastocapillarity; those that do have focused on a liquid dewetting off a bulk elastic solid and have chiefly been theoretical in nature~\cite{Martin1998,Martin2000,Martin2001,Gerardin2005,Qiao2008}. Some studies have investigated systems where the liquid film is capped by a thin, compliant elastic layer~\cite{Becker2002,Kumar2004,deBeer2009,Carlson2015}. In one set of experiments, the dewetting of thin water films between two sheets of mica, several microns thick, was observed~\cite{Becker2002,deBeer2009}. Although the effect of changing the stiffness of the elastic layer was not systematically studied, it was noted that thicker  mica sheets (less compliant) resulted in slower dewetting. A similar result has been recovered in a recent theoretical study~\cite{Carlson2015}.

Here, we study the dewetting of a liquid film which is capped by a taut, thin elastic film. Increasing tension in the elastic layer leads to a reduction in the hole growth rate, a flatter and wider dewetting rim, as well as increased stability of the rim. In addition, we show for the first time that elastic capping films can be utilized to generate novel dewetting morphologies. A biaxial tension causes holes to appear with an elongated, non-circular shape. Further, we show that these elastic boundaries can be manipulated to generate holes with a square morphology.

\begin{figure}[t!]
     \includegraphics[width=8.89cm]{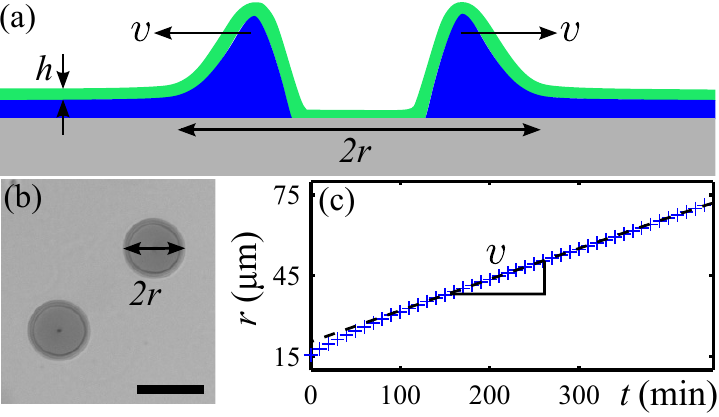}
\caption{(a) Schematic of PS film dewetting atop silicon and capped by a thin elastomeric film. The material from the hole collects in a rim which moves outward with speed $v$. (b) Optical micrograph of two holes capped by an elastomeric film with an isotropic strain $\epsilon =$~28\% after 250 min. Scale bar = 100 $\mu$m. (c) Radius of a hole in the experiment depicted in (b) over time. The dashed line is the best fit to the linear growth region, and its slope is the dewetting speed $v$. }
\label{fig1}
\end{figure}

In these experiments, thin polystyrene (PS) films of thickness $h_\mathrm{PS} \sim 100$~nm (Scientific Polymer Products, number averaged molecular weight $M_n$ = 15,800 g/mol, polydispersity index 1.05) are prepared through spincoating out of a toluene solution onto $10 \mathrm{\ mm} \times 10 \mathrm{\ mm}$ silicon substrates. The PS is removed near the four edges of the silicon using an acetone-wetted cotton swab. The samples are annealed above the glass transition of PS ($T_g\sim 100^\circ$C) at 140$^\circ$C for 10 min to relax the polymer chains and remove any residual solvent. Free-standing elastomeric films ranging in thickness from 50 -- 320 nm are prepared from Elastollan TPU 1185A (BASF) by spincoating onto freshly cleaved mica substrates and transferred onto a home-built straining set-up~\cite{SI}. The Elastollan films are stretched isotropically with strain $\epsilon$, or biaxially with strains $\epsilon_\mathrm{low}$ and $\epsilon_\mathrm{high}$ in two orthogonal directions. The initial film thickness of the Elastollan is chosen such that the final film thickness after straining, $h$, is a fixed quantity. In this study, we test $h = 50$~nm and 100~nm. The strained films are then transferred onto the PS sample.  Having removed PS from the edges of the silicon wafer, the Elastollan makes good contact with the silicon around the perimeter of the sample, which ensures that the pre-strain in the film cannot relax. To observe dewetting of the PS, samples are annealed at 140$^\circ$C. Within minutes of heating, small holes form in the PS film (through heterogeneous nucleation) surrounded by rims where the liquid has collected, as depicted schematically in Fig.~\ref{fig1}(a). These holes quickly adopt a circular shape and the elastomeric film acts as a solid capping layer during this process. The growth of several holes is monitored using an optical microscope (Fig.~\ref{fig1}(b)) and the radius of each hole, $r$, is measured over time (Fig.~\ref{fig1}(c)). For the first $\sim 100$~min, the hole growth speed is changing with time, an effect which may be attributed to complex dynamics or liquid slip at the contact line being important at early times~\cite{Jacobs1998,Ghatak1999,Munch2005}. Thereafter, the speed of dewetting, $v$,  tends towards a constant, consistent with standard dewetting (i.e. Newtonian, viscous, non-slipping liquid on a rigid surface)~\cite{Redon1991,Gennes2008}. The holes are tracked until the rim exhibits significant morphological changes due to undergoing an instability akin to the Plateau-Rayleigh instability (PRI)~\cite{Sharma1996,Choi2006,Baumchen2014}.  

\begin{figure}[t!]
     \includegraphics[width=8.89cm]{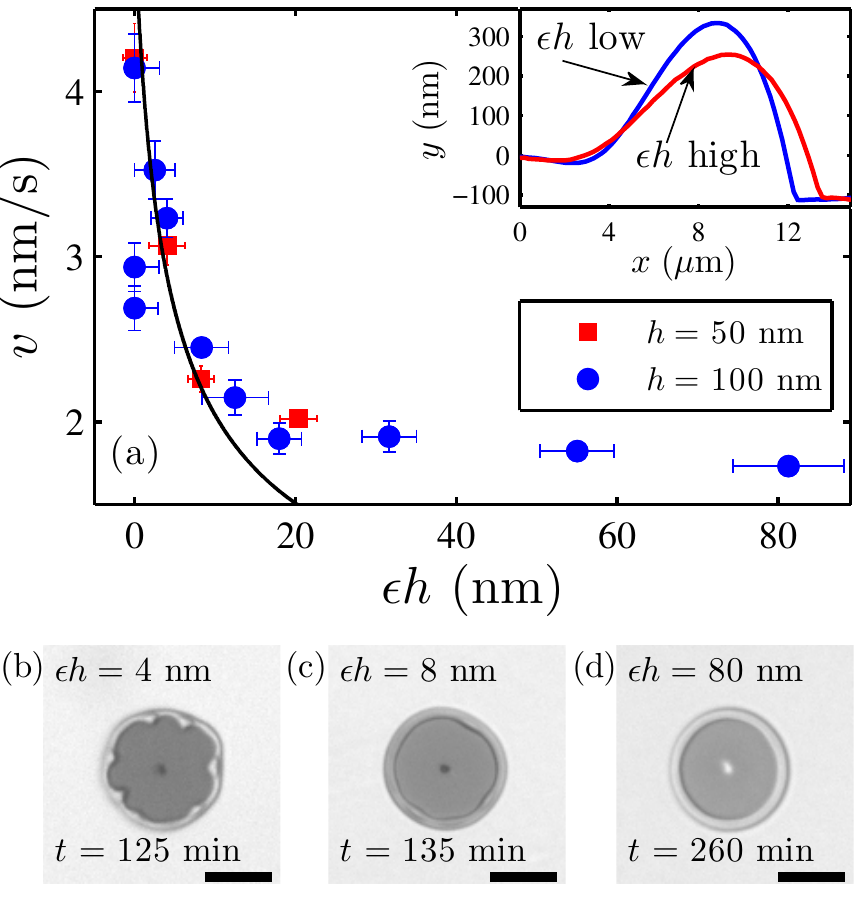}
\caption{(a) Dewetting speed as a function of $\epsilon h$, which is proportional to the tension, for two different elastomeric film thicknesses. The solid curve represents the the minimal theoretical model for $v(\epsilon h)$ described in the text. Inset: AFM profiles of the rim shape for a low ($\epsilon h$ =~4 nm) and high ($\epsilon h$ =~80 nm) tension sample. (b)-(d) Optical micrographs of holes of the same size but with different tension in the capping elastomer. Scale bars = 50 $\mu$m.}
\label{fig2}
\end{figure}

In the first part of this study, we investigate how the dewetting speed is influenced by changes in the isotropic strain of the capping elastomer. Since we are changing the strain in the films while keeping their final thickness constant, we are primarily changing the tension, $T$, in the films, while maintaining a nearly constant bending rigidity. In the case of a Hookean material, the tension is simply related to the strain, $T \propto \epsilon h$~\cite{SI}. Although Elastollan is not Hookean over large strains~\cite{elastollan_modulus3}, $\epsilon h$ is an approximate indicator of the tension in the film. Thus, in Fig.~\ref{fig2}(a), we plot the dewetting speed $v$ as a function of $\epsilon h$ for two different values of $h$. We see  that the dewetting speed decreases with increasing strain in the elastomer. In addition, changing the elastomer thickness by a factor of two (i.e. an eight-fold change in the bending rigidity) produces no observable change in $v$. This observation implies that bending of the elastomer does not determine the dewetting rate, but rather the tension in the film plays the dominant role. In the inset of Fig.~\ref{fig2}(a), we show atomic force microscopy (AFM) profiles of rims for samples with $\epsilon h \sim$~4 nm and $\epsilon h \sim$~80 nm. From these scans, it is clear that increased tension results in a flatter, wider rim. 

In the simplest model, we assume that the tension is not globally altered by the formation of holes. This simplistic picture is consistent with the  observations that the dewetting rate (which depends on $T$) is constant as holes grow and independent of the number of nucleated holes on a sample. Within this approximation, the energy per unit area of the wet regions of the sample is $E_\mathrm{wet} = T+\gamma_\mathrm{el,v}+\gamma_\mathrm{el,l}+\gamma_\mathrm{s,l}$, where $\gamma$ represents interfacial tensions, ``el", ``v", ``l" and ``s" denote the elastomer, vapour, liquid and the solid substrate. The energy per unit area of the dry regions is $E_\mathrm{dry} = T+\gamma_\mathrm{el,v}+\gamma_\mathrm{el,s}$. Therefore,  the spreading parameter $S = E_\mathrm{dry} - E_\mathrm{wet}$ is not dependent on the tension, and thus, the driving force for dewetting remains unchanged when the tension is altered.  Thus, the change in tension must alter the dissipation in the system. 
A simple balance of mechanical and interfacial tensions at the contact line (see~\cite{Schulman2016,Schulman2017} and the supplemental information~\cite{SI}), yields a relation for the equilibrium contact angle $\theta_\mathrm{E} \approx \theta_0/\sqrt{1+T/(\gamma_\mathrm{el,v}+\gamma_\mathrm{el,l})}$ of the capped rim in the limit of small angles, where $\theta_0$ is the equilibrium contact angle for vanishing mechanical tension. This expression is qualitatively consistent with the inset of Fig.~\ref{fig2}(a), where an increased tension results in a flatter rim. If the dissipation is dominated by viscous dissipation at the contact line, then $v \propto \theta_\mathrm{D}$, where $\theta_\mathrm{D}$ is the dynamic contact angle of the capped rim~\cite{Gennes2008}. Assuming that $\theta_\mathrm{D} \approx \theta_\mathrm{E}$, we find $v \approx v_0/\sqrt{1+T/(\gamma_\mathrm{el,v}+\gamma_\mathrm{el,l})}$, where $v_0 $ is the dewetting velocity with vanishing tension. To facilitate a quantitative comparison with the data, we make the simplifying assumption of Hookean elasticity such that $T = 2E\epsilon h$ with $E \sim 10^7$ Pa~\cite{SI,elastollan_modulus3}, and to plot $v(\epsilon h)$ we choose reasonable values for the free parameters: $\gamma_\mathrm{el,v}+\gamma_\mathrm{el,l} \sim$~40 mJ/m$^2$ and $v_0 \sim$~5 nm/s. As seen in Fig.~\ref{fig2}(a), only qualitative agreement between $v(T)$ and the data is seen, which is expected given the assumptions and approximations outlined above, including the fact that the role played by bending at the contact line has been neglected. If hydrodynamic slip is important in this system, the dissipation may be area-dependent rather than concentrated at the contact line; but for either of these mechanisms, a flatter and wider rim will result in increased dissipation~\cite{Redon1991,Redon1994,Jacobs1998}. We observe a constant dewetting speed at large times, indicating that contact line dissipation likely dominates at these times.

Changing tension also affects the stability of the liquid rim, as seen in Figs.~\ref{fig2}(b)-(d). Despite the holes being equal in size, the low tension sample shows rims which have reached late stages of the rim instability as fingers are in the process of forming, the intermediate sample shows bulges in the rims, while the high tension sample exhibits rims which appear unaffected by the instability. It is known that liquids dewetting off more wettable substrates (i.e. lower dynamic contact angle, wider rims) are less susceptible to developing the rim instability~\cite{Sharma1996,Choi2006}. Analogously, here the higher tension samples are characterized by rims which are flatter and wider, and this leads to increased rim stability (consistent with the PRI: a lower curvature increases stability). 

\begin{figure}[h]
     \includegraphics[width=8.89cm]{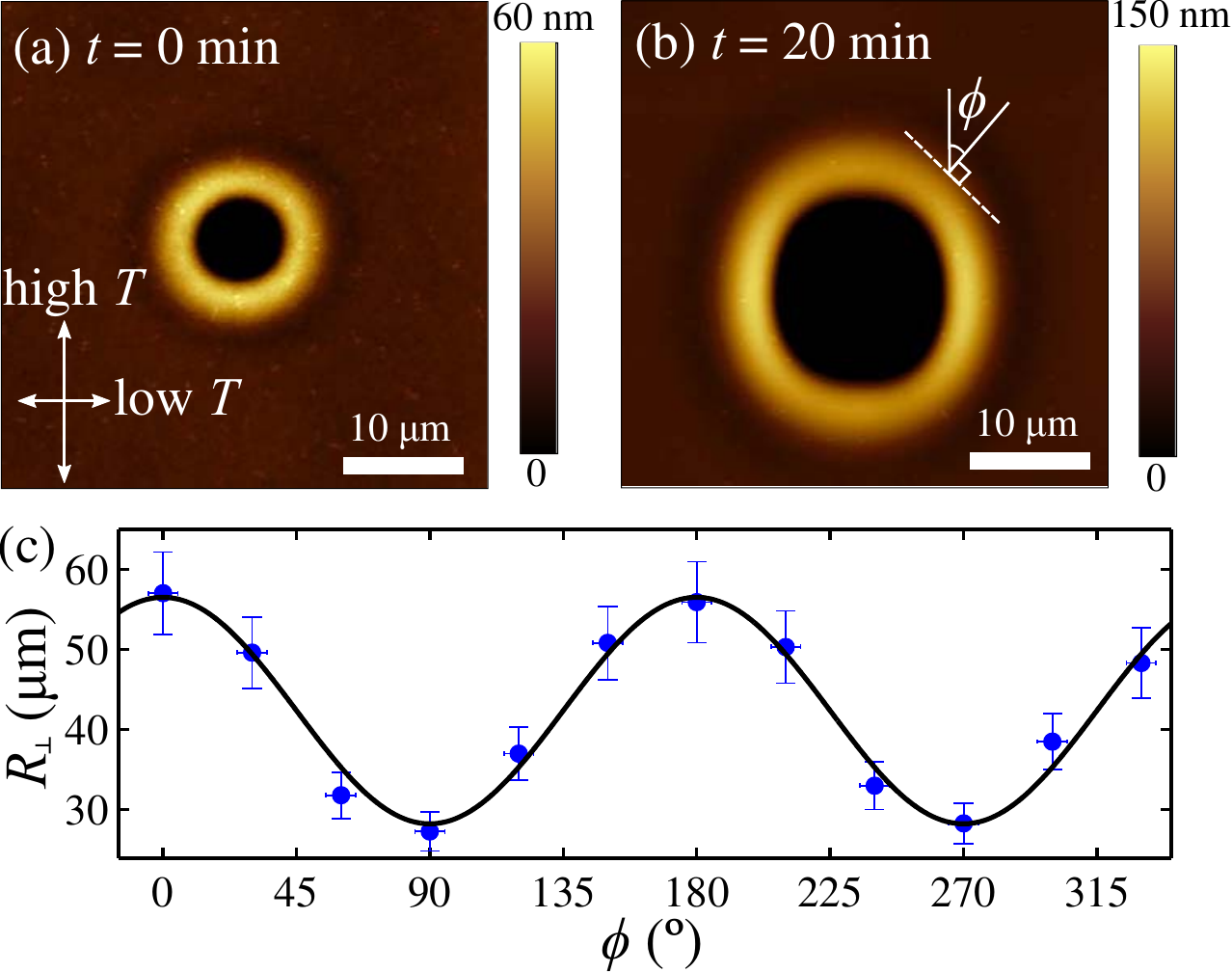}
\caption{(a) AFM scan of the initially round hole which has been capped by an elastomeric film with $\epsilon_\mathrm{high} \sim 100$~\%, $\epsilon_\mathrm{low} \sim$~0\% and $h\sim$~100 nm. (b) AFM scan of the same hole after annealing. (c) For the hole at the time of (b), $R_\perp$ as a function of the orientation of the rim (defined in (b)) The solid curve represents the best fit of the expression for  $R_\perp ( \phi)$ discussed in the text ($R_0 = 28~\mu$m).  In the AFM scans, colours indicate relative heights on the sample, and the bottom of the hole is outside of the colour bar range. The high and low tension axes are oriented as shown in (a).}
\label{fig3}
\end{figure}

\begin{figure*}[t!]
     \includegraphics[width=18cm]{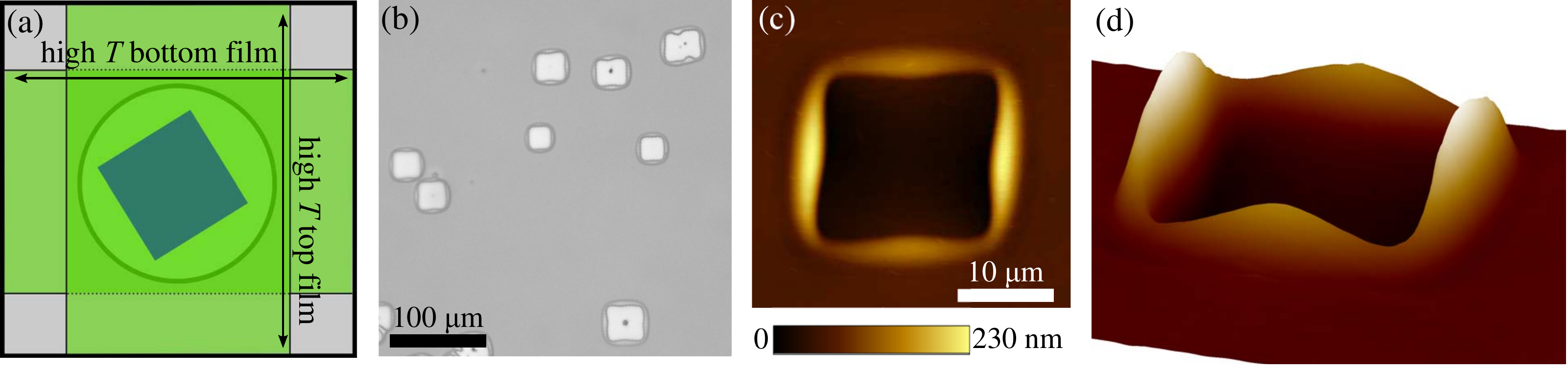}
\caption{(a) Top view schematic of the sandwiched sample. The trilayer sample is free-standing over the hole in the washer, and the PS is sandwiched between two biaxially stretched elastic films. (b) Optical micrograph of the square holes nucleating in the PS. (c-d) AFM scan of a square hole. Colours indicate relative heights on the sample.}
\label{fig4}
\end{figure*}

There is great interest in utilizing dewetting for microscopic pattern formation~\cite{Rockford1999,Higgins2000,Konnur2000,Kargupta2002,Suh2002,Sehgal2002}. Although a theoretical study has shown that spinodal dewetting patterns become anisotropic atop biaxially strained bulk elastic substrates~\cite{Qiao2008}, there has been no subsequent work studying the possibility of exploiting elasticity to generate novel morphologies. Thus, we perform the first investigation of how hole morphology may be altered using a capping elastomeric layer. In these experiments, we anneal the \emph{uncapped} PS samples long enough that circular holes form.  An elastic film is strained only along one direction and held fixed in the perpendicular direction ($\epsilon_\mathrm{high}h \sim$~100 nm, $\epsilon_\mathrm{low}h \sim$~ 0) and then transferred onto the sample, and the experiment proceeds as usual. As seen in Fig.~\ref{fig3}(a), an AFM scan reveals that the initial hole shape is completely circular, as we have prepared it. However, after a short time, the hole adopts a shape which is elongated along the high tension direction (Fig.~\ref{fig3}(b)). Holes that form after capping are also elongated in the same way, but we purposefully choose to start with a small circular hole, to ensure a more robust initial condition. At first, the elongated hole is surprising because Fig.~\ref{fig2}(a) implies that holes grow slower with high tension, and na\"{i}vely, we would thereby expect holes to be elongated along the low tension direction. However, since we are investigating the {\it initial} shape of the hole here, it necessarily implies that the energetic cost of forming the rim determines the morphology, and the physics pertaining to the later stages of dewetting, where the growth rate of holes was constant in time, is not applicable (see supplemental information for further discussion~\cite{SI}). 
To create the rim, extra interface is created, and work is done against mechanical and interfacial tensions. It is then favourable to deform the film less across the high tension direction and  more across the low tension direction -- this balance leads to an elongated shape shown in Fig.~\ref{fig3}(b). This argument is analogous to deforming a trampoline with an anisotropic tension -- an asymmetric deformation would result.

Consistent with the case of isotropic tension, the elongated hole (Fig.~\ref{fig3}(b)) has a flatter rim along the high tension direction. We quantify the rim's shape by the radius of curvature of the top of the rim for a slice normal to the hole , $R_\perp$ (i.e. the radius of curvature at the peak of the rim profiles akin to those shown in the inset of Fig.~\ref{fig2}(a)). The radius of curvature of the rim for a slice in the tangential direction is much larger and need not be considered. For the hole shown in Fig.~\ref{fig3}(b), we measure $R_\perp$ for different values of $\phi$, defined as the angle the normal of the rim subtends to the high tension direction, indicated in Fig.~\ref{fig3}(b). These data are plotted in Fig.~\ref{fig3}(c), where it is clear that the radius of curvature of the rim is much larger along the high tension direction. At these early times, wherein the rim energetics dictate the hole's shape, we expect that the liquid within the entire rim has a constant Laplace pressure. If bending is ignored, as validated above, the pressure in the rim is $P \approx T_\mathrm{tot}/R_\perp$, where $T_\mathrm{tot}$ is the total mechanical and interfacial tensions between the elastomer-air interface and elastomer-liquid interface. However, at these large strains, the mechanical tension is dominant over interfacial tensions, since the Young's modulus of Elastollan is $\sim 10^7$ Pa~\cite{elastollan_modulus3,SI}. Thus, we can simply write $R_\perp \approx P T$. Since the film is prepared with a biaxial strain, $T = T_\mathrm{high} \mathrm{cos}^2\phi + T_\mathrm{low} \mathrm{sin}^2\phi$. For a Hookean material with a Poisson's ratio of 0.5 (typical of elastomers), it is the case that $T_\mathrm{high} = 2 T_\mathrm{low}$ when $\epsilon_\mathrm{low} = 0$. We can use this fact to generate an {\it approximate} expression for $R_\perp$, and find that
$R_\perp  \approx  R_0(2 \mathrm{cos}^2\phi + \mathrm{sin}^2\phi)$,
where $R_0$ is a constant. In Fig.~\ref{fig3}(c), the solid curve represents the best fit of the expression for $R_\perp ( \phi)$, with $R_0$ as the only free-parameter. Despite the approximations made,  $R_\perp ( \phi)$ captures the curvature of the rim well.

In the experiments described thus far, the PS film is sandwiched between an elastic film and a rigid silicon substrate. However, the rigid substrate may also be substituted for an elastic film.  To further manipulate the boundary conditions for dewetting, we make the sample depicted in Fig.~\ref{fig4}(a). Here, a biaxially strained elastomeric film ($\epsilon_\mathrm{high}h \sim$~ 140 nm) is transferred onto a steel washer with a circular hole. The PS film is placed atop this elastomer. Finally, a second elastomeric film (strained biaxially in the same way) is placed on top, but with its high tension direction oriented perpendicular to that of the bottom elastomeric film. As such, the sample is a free-standing trilayer: elastomer-liquid-elastomer. When this sample is  annealed, the holes evolve with a  \emph{square morphology} with the sides oriented along the principal strain directions. This configuration is maintained as the holes continue to grow (Fig.~\ref{fig4}(b), see video \cite{SI}). As shown in Fig.~\ref{fig4}(c)-(d), an AFM scan of one of these holes performed from the top side of this sample reveals that the rims parallel to the high tension direction of the top film are much taller than the ones parallel to the low tension direction. On the underside of the sample where the high tension direction in the film is perpendicular to that of the top film, we would expect the 3D-morphology in Fig.~\ref{fig4}(c)-(d) to be rotated by 90$^\circ$. This morphology is completely consistent with the notion that the rim energetics dictate the initial shape of the hole.  Precisely as in the elongated hole in Fig.~\ref{fig3}(b), it is favourable to minimize the deformation of the elastic film across the high tension direction and maximize it across the low tension direction.

In this study, we have investigated the dewetting of a thin liquid film capped by a taut elastomeric layer with both isotropic and biaxial tension. For the case of isotropic tension, holes are  round and dewet at a constant speed at long times. A higher tension in the elastomer leads to a flattened rim, which increases the dissipation and thus decreases the dewetting speed. When the capping elastic film is prepared with a biaxial tension, holes assume a shape which is elongated along the high tension direction, a process driven by the energetics of the rim. By choosing the magnitude and anisotropy of the tension in the elastic film, holes can be designed to have wider and more stable rims or to form with non-circular shapes. In addition, when a liquid film is sandwiched between two elastomeric layers with biaxial tension, holes acquire a square morphology. We have shown that by using thin elastic films to cap the dewetting liquid, a new avenue for patterning emerges.

\begin{acknowledgments}
The authors thank Joshua D. McGraw, Elie Rapha\"el, and Thomas Salez for valuable discussions. The financial support by the Natural Science and Engineering Research Council of Canada is gratefully acknowledged.
\end{acknowledgments}


\begin{thebibliography}{47}%
\makeatletter
\providecommand \@ifxundefined [1]{%
 \@ifx{#1\undefined}
}%
\providecommand \@ifnum [1]{%
 \ifnum #1\expandafter \@firstoftwo
 \else \expandafter \@secondoftwo
 \fi
}%
\providecommand \@ifx [1]{%
 \ifx #1\expandafter \@firstoftwo
 \else \expandafter \@secondoftwo
 \fi
}%
\providecommand \natexlab [1]{#1}%
\providecommand \enquote  [1]{``#1''}%
\providecommand \bibnamefont  [1]{#1}%
\providecommand \bibfnamefont [1]{#1}%
\providecommand \citenamefont [1]{#1}%
\providecommand \href@noop [0]{\@secondoftwo}%
\providecommand \href [0]{\begingroup \@sanitize@url \@href}%
\providecommand \@href[1]{\@@startlink{#1}\@@href}%
\providecommand \@@href[1]{\endgroup#1\@@endlink}%
\providecommand \@sanitize@url [0]{\catcode `\\12\catcode `\$12\catcode
  `\&12\catcode `\#12\catcode `\^12\catcode `\_12\catcode `\%12\relax}%
\providecommand \@@startlink[1]{}%
\providecommand \@@endlink[0]{}%
\providecommand \url  [0]{\begingroup\@sanitize@url \@url }%
\providecommand \@url [1]{\endgroup\@href {#1}{\urlprefix }}%
\providecommand \urlprefix  [0]{URL }%
\providecommand \Eprint [0]{\href }%
\providecommand \doibase [0]{http://dx.doi.org/}%
\providecommand \selectlanguage [0]{\@gobble}%
\providecommand \bibinfo  [0]{\@secondoftwo}%
\providecommand \bibfield  [0]{\@secondoftwo}%
\providecommand \translation [1]{[#1]}%
\providecommand \BibitemOpen [0]{}%
\providecommand \bibitemStop [0]{}%
\providecommand \bibitemNoStop [0]{.\EOS\space}%
\providecommand \EOS [0]{\spacefactor3000\relax}%
\providecommand \BibitemShut  [1]{\csname bibitem#1\endcsname}%
\let\auto@bib@innerbib\@empty
\bibitem [{\citenamefont {de~Gennes}\ \emph {et~al.}(2008)\citenamefont
  {de~Gennes}, \citenamefont {Brochard-Wyart},\ and\ \citenamefont
  {Quere}}]{Gennes2008}%
  \BibitemOpen
  \bibfield  {author} {\bibinfo {author} {\bibfnamefont {P.}~\bibnamefont
  {de~Gennes}}, \bibinfo {author} {\bibfnamefont {F.}~\bibnamefont
  {Brochard-Wyart}}, \ and\ \bibinfo {author} {\bibfnamefont {D.}~\bibnamefont
  {Quere}},\ }\href {http://www.ulb.tu-darmstadt.de/tocs/114142769.pdf} {\emph
  {\bibinfo {title} {{Capillarity and Wetting Phenomena}}}}\ (\bibinfo
  {publisher} {Springer},\ \bibinfo {year} {2008})\BibitemShut {NoStop}%
\bibitem [{\citenamefont {Redon}\ \emph {et~al.}(1991)\citenamefont {Redon},
  \citenamefont {Brochard-Wyart},\ and\ \citenamefont {Rondelez}}]{Redon1991}%
  \BibitemOpen
  \bibfield  {author} {\bibinfo {author} {\bibfnamefont {C.}~\bibnamefont
  {Redon}}, \bibinfo {author} {\bibfnamefont {F.}~\bibnamefont
  {Brochard-Wyart}}, \ and\ \bibinfo {author} {\bibfnamefont {F.}~\bibnamefont
  {Rondelez}},\ }\href@noop {} {\bibfield  {journal} {\bibinfo  {journal}
  {Phys. Rev. Lett.}\ }\textbf {\bibinfo {volume} {66}},\ \bibinfo {pages}
  {715} (\bibinfo {year} {1991})}\BibitemShut {NoStop}%
\bibitem [{\citenamefont {Reiter}(1992)}]{Reiter1992}%
  \BibitemOpen
  \bibfield  {author} {\bibinfo {author} {\bibfnamefont {G.}~\bibnamefont
  {Reiter}},\ }\href@noop {} {\bibfield  {journal} {\bibinfo  {journal} {Phys.
  Rev. Lett.}\ }\textbf {\bibinfo {volume} {68}},\ \bibinfo {pages} {75}
  (\bibinfo {year} {1992})}\BibitemShut {NoStop}%
\bibitem [{\citenamefont {Reiter}\ \emph {et~al.}(2005)\citenamefont {Reiter},
  \citenamefont {Hamieh}, \citenamefont {Damman}, \citenamefont {Sclavons},
  \citenamefont {Sylvain}, \citenamefont {Vilmin},\ and\ \citenamefont
  {Rapha{\"e}l}}]{Reiter2005}%
  \BibitemOpen
  \bibfield  {author} {\bibinfo {author} {\bibfnamefont {G.}~\bibnamefont
  {Reiter}}, \bibinfo {author} {\bibfnamefont {M.}~\bibnamefont {Hamieh}},
  \bibinfo {author} {\bibfnamefont {P.}~\bibnamefont {Damman}}, \bibinfo
  {author} {\bibfnamefont {S.}~\bibnamefont {Sclavons}}, \bibinfo {author}
  {\bibfnamefont {G.}~\bibnamefont {Sylvain}}, \bibinfo {author} {\bibfnamefont
  {T.}~\bibnamefont {Vilmin}}, \ and\ \bibinfo {author} {\bibfnamefont
  {E.}~\bibnamefont {Rapha{\"e}l}},\ }\href@noop {} {\bibfield  {journal}
  {\bibinfo  {journal} {Nat. Mater.}\ }\textbf {\bibinfo {volume} {4}},\
  \bibinfo {pages} {754} (\bibinfo {year} {2005})}\BibitemShut {NoStop}%
\bibitem [{\citenamefont {Damman}\ \emph {et~al.}(2007)\citenamefont {Damman},
  \citenamefont {Gabriele}, \citenamefont {Copp\'ee}, \citenamefont {Desprez},
  \citenamefont {Villers}, \citenamefont {Vilmin}, \citenamefont {Rapha\"el},
  \citenamefont {Hamieh}, \citenamefont {Akhrass},\ and\ \citenamefont
  {Reiter}}]{Damman2007}%
  \BibitemOpen
  \bibfield  {author} {\bibinfo {author} {\bibfnamefont {P.}~\bibnamefont
  {Damman}}, \bibinfo {author} {\bibfnamefont {S.}~\bibnamefont {Gabriele}},
  \bibinfo {author} {\bibfnamefont {S.}~\bibnamefont {Copp\'ee}}, \bibinfo
  {author} {\bibfnamefont {S.}~\bibnamefont {Desprez}}, \bibinfo {author}
  {\bibfnamefont {D.}~\bibnamefont {Villers}}, \bibinfo {author} {\bibfnamefont
  {T.}~\bibnamefont {Vilmin}}, \bibinfo {author} {\bibfnamefont
  {E.}~\bibnamefont {Rapha\"el}}, \bibinfo {author} {\bibfnamefont
  {M.}~\bibnamefont {Hamieh}}, \bibinfo {author} {\bibfnamefont {S.~A.}\
  \bibnamefont {Akhrass}}, \ and\ \bibinfo {author} {\bibfnamefont
  {G.}~\bibnamefont {Reiter}},\ }\href {\doibase 10.1103/PhysRevLett.99.036101}
  {\bibfield  {journal} {\bibinfo  {journal} {Phys. Rev. Lett.}\ }\textbf
  {\bibinfo {volume} {99}},\ \bibinfo {pages} {036101} (\bibinfo {year}
  {2007})}\BibitemShut {NoStop}%
\bibitem [{\citenamefont {Dalnoki-Veress}\ \emph {et~al.}(1999)\citenamefont
  {Dalnoki-Veress}, \citenamefont {Nickel}, \citenamefont {Roth},\ and\
  \citenamefont {Dutcher}}]{Dalnoki1999}%
  \BibitemOpen
  \bibfield  {author} {\bibinfo {author} {\bibfnamefont {K.}~\bibnamefont
  {Dalnoki-Veress}}, \bibinfo {author} {\bibfnamefont {B.}~\bibnamefont
  {Nickel}}, \bibinfo {author} {\bibfnamefont {C.}~\bibnamefont {Roth}}, \ and\
  \bibinfo {author} {\bibfnamefont {J.}~\bibnamefont {Dutcher}},\ }\href@noop
  {} {\bibfield  {journal} {\bibinfo  {journal} {Phys. Rev. E}\ }\textbf
  {\bibinfo {volume} {59}},\ \bibinfo {pages} {2153} (\bibinfo {year}
  {1999})}\BibitemShut {NoStop}%
\bibitem [{\citenamefont {Herminghaus}\ \emph {et~al.}(2002)\citenamefont
  {Herminghaus}, \citenamefont {Seemann},\ and\ \citenamefont
  {Jacobs}}]{Herminghaus2002}%
  \BibitemOpen
  \bibfield  {author} {\bibinfo {author} {\bibfnamefont {S.}~\bibnamefont
  {Herminghaus}}, \bibinfo {author} {\bibfnamefont {R.}~\bibnamefont
  {Seemann}}, \ and\ \bibinfo {author} {\bibfnamefont {K.}~\bibnamefont
  {Jacobs}},\ }\href {\doibase 10.1103/PhysRevLett.89.056101} {\bibfield
  {journal} {\bibinfo  {journal} {Phys. Rev. Lett.}\ }\textbf {\bibinfo
  {volume} {89}},\ \bibinfo {pages} {056101} (\bibinfo {year}
  {2002})}\BibitemShut {NoStop}%
\bibitem [{\citenamefont {Gabriele}\ \emph {et~al.}(2006)\citenamefont
  {Gabriele}, \citenamefont {Sclavons}, \citenamefont {Reiter},\ and\
  \citenamefont {Damman}}]{Gabriele2006}%
  \BibitemOpen
  \bibfield  {author} {\bibinfo {author} {\bibfnamefont {S.}~\bibnamefont
  {Gabriele}}, \bibinfo {author} {\bibfnamefont {S.}~\bibnamefont {Sclavons}},
  \bibinfo {author} {\bibfnamefont {G.}~\bibnamefont {Reiter}}, \ and\ \bibinfo
  {author} {\bibfnamefont {P.}~\bibnamefont {Damman}},\ }\href@noop {}
  {\bibfield  {journal} {\bibinfo  {journal} {Phys. Rev. Lett.}\ }\textbf
  {\bibinfo {volume} {96}},\ \bibinfo {pages} {156105} (\bibinfo {year}
  {2006})}\BibitemShut {NoStop}%
\bibitem [{\citenamefont {Redon}\ \emph {et~al.}(1994)\citenamefont {Redon},
  \citenamefont {Brzoska},\ and\ \citenamefont {Brochard-Wyart}}]{Redon1994}%
  \BibitemOpen
  \bibfield  {author} {\bibinfo {author} {\bibfnamefont {C.}~\bibnamefont
  {Redon}}, \bibinfo {author} {\bibfnamefont {J.}~\bibnamefont {Brzoska}}, \
  and\ \bibinfo {author} {\bibfnamefont {F.}~\bibnamefont {Brochard-Wyart}},\
  }\href@noop {} {\bibfield  {journal} {\bibinfo  {journal} {Macromolecules}\
  }\textbf {\bibinfo {volume} {27}},\ \bibinfo {pages} {468} (\bibinfo {year}
  {1994})}\BibitemShut {NoStop}%
\bibitem [{\citenamefont {Fetzer}\ \emph {et~al.}(2005)\citenamefont {Fetzer},
  \citenamefont {Jacobs}, \citenamefont {M{\"u}nch}, \citenamefont {Wagner},\
  and\ \citenamefont {Witelski}}]{Fetzer2005}%
  \BibitemOpen
  \bibfield  {author} {\bibinfo {author} {\bibfnamefont {R.}~\bibnamefont
  {Fetzer}}, \bibinfo {author} {\bibfnamefont {K.}~\bibnamefont {Jacobs}},
  \bibinfo {author} {\bibfnamefont {A.}~\bibnamefont {M{\"u}nch}}, \bibinfo
  {author} {\bibfnamefont {B.}~\bibnamefont {Wagner}}, \ and\ \bibinfo {author}
  {\bibfnamefont {T.}~\bibnamefont {Witelski}},\ }\href@noop {} {\bibfield
  {journal} {\bibinfo  {journal} {Phys. Rev. Lett.}\ }\textbf {\bibinfo
  {volume} {95}},\ \bibinfo {pages} {127801} (\bibinfo {year}
  {2005})}\BibitemShut {NoStop}%
\bibitem [{\citenamefont {B\"{a}umchen}\ \emph {et~al.}(2010)\citenamefont
  {B\"{a}umchen}, \citenamefont {Lessel}, \citenamefont {Fetzer}, \citenamefont
  {Seemann},\ and\ \citenamefont {Jacobs}}]{Baumchen2010}%
  \BibitemOpen
  \bibfield  {author} {\bibinfo {author} {\bibfnamefont {O.}~\bibnamefont
  {B\"{a}umchen}}, \bibinfo {author} {\bibfnamefont {M.}~\bibnamefont
  {Lessel}}, \bibinfo {author} {\bibfnamefont {R.}~\bibnamefont {Fetzer}},
  \bibinfo {author} {\bibfnamefont {R.}~\bibnamefont {Seemann}}, \ and\
  \bibinfo {author} {\bibfnamefont {K.}~\bibnamefont {Jacobs}},\ }\href@noop {}
  {\bibfield  {journal} {\bibinfo  {journal} {J. Phys. Conf. Ser.}\ }\textbf
  {\bibinfo {volume} {216}},\ \bibinfo {pages} {012002} (\bibinfo {year}
  {2010})}\BibitemShut {NoStop}%
\bibitem [{\citenamefont {McGraw}\ \emph {et~al.}(2016)\citenamefont {McGraw},
  \citenamefont {Chan}, \citenamefont {Maurer}, \citenamefont {Salez},
  \citenamefont {Benzaquen}, \citenamefont {Rapha\"{e}l}, \citenamefont
  {Brinkmann},\ and\ \citenamefont {Jacobs}}]{McGraw2016}%
  \BibitemOpen
  \bibfield  {author} {\bibinfo {author} {\bibfnamefont {J.~D.}\ \bibnamefont
  {McGraw}}, \bibinfo {author} {\bibfnamefont {T.~S.}\ \bibnamefont {Chan}},
  \bibinfo {author} {\bibfnamefont {S.}~\bibnamefont {Maurer}}, \bibinfo
  {author} {\bibfnamefont {T.}~\bibnamefont {Salez}}, \bibinfo {author}
  {\bibfnamefont {M.}~\bibnamefont {Benzaquen}}, \bibinfo {author}
  {\bibfnamefont {E.}~\bibnamefont {Rapha\"{e}l}}, \bibinfo {author}
  {\bibfnamefont {M.}~\bibnamefont {Brinkmann}}, \ and\ \bibinfo {author}
  {\bibfnamefont {K.}~\bibnamefont {Jacobs}},\ }\href {\doibase
  10.1073/pnas.1513565113} {\bibfield  {journal} {\bibinfo  {journal} {Proc.
  Natl. Acad. Sci. U.S.A.}\ }\textbf {\bibinfo {volume} {113}},\ \bibinfo
  {pages} {1168} (\bibinfo {year} {2016})}\BibitemShut {NoStop}%
\bibitem [{\citenamefont {Seemann}\ \emph
  {et~al.}(2001{\natexlab{a}})\citenamefont {Seemann}, \citenamefont
  {Herminghaus},\ and\ \citenamefont {Jacobs}}]{Seemann2001}%
  \BibitemOpen
  \bibfield  {author} {\bibinfo {author} {\bibfnamefont {R.}~\bibnamefont
  {Seemann}}, \bibinfo {author} {\bibfnamefont {S.}~\bibnamefont
  {Herminghaus}}, \ and\ \bibinfo {author} {\bibfnamefont {K.}~\bibnamefont
  {Jacobs}},\ }\href@noop {} {\bibfield  {journal} {\bibinfo  {journal} {Phys.
  Rev. Lett.}\ }\textbf {\bibinfo {volume} {86}},\ \bibinfo {pages} {5534}
  (\bibinfo {year} {2001}{\natexlab{a}})}\BibitemShut {NoStop}%
\bibitem [{\citenamefont {Seemann}\ \emph
  {et~al.}(2001{\natexlab{b}})\citenamefont {Seemann}, \citenamefont
  {Herminghaus},\ and\ \citenamefont {Jacobs}}]{Seemann2001b}%
  \BibitemOpen
  \bibfield  {author} {\bibinfo {author} {\bibfnamefont {R.}~\bibnamefont
  {Seemann}}, \bibinfo {author} {\bibfnamefont {S.}~\bibnamefont
  {Herminghaus}}, \ and\ \bibinfo {author} {\bibfnamefont {K.}~\bibnamefont
  {Jacobs}},\ }\href@noop {} {\bibfield  {journal} {\bibinfo  {journal} {J.
  Phys. Cond. Mat.}\ }\textbf {\bibinfo {volume} {13}},\ \bibinfo {pages}
  {4925} (\bibinfo {year} {2001}{\natexlab{b}})}\BibitemShut {NoStop}%
\bibitem [{\citenamefont {De~Silva}\ \emph {et~al.}(2007)\citenamefont
  {De~Silva}, \citenamefont {Geoghegan}, \citenamefont {Higgins}, \citenamefont
  {Krausch}, \citenamefont {David},\ and\ \citenamefont
  {Reiter}}]{DeSilva2007}%
  \BibitemOpen
  \bibfield  {author} {\bibinfo {author} {\bibfnamefont {J.}~\bibnamefont
  {De~Silva}}, \bibinfo {author} {\bibfnamefont {M.}~\bibnamefont {Geoghegan}},
  \bibinfo {author} {\bibfnamefont {A.}~\bibnamefont {Higgins}}, \bibinfo
  {author} {\bibfnamefont {G.}~\bibnamefont {Krausch}}, \bibinfo {author}
  {\bibfnamefont {M.-O.}\ \bibnamefont {David}}, \ and\ \bibinfo {author}
  {\bibfnamefont {G.}~\bibnamefont {Reiter}},\ }\href@noop {} {\bibfield
  {journal} {\bibinfo  {journal} {Phys. Rev. Lett.}\ }\textbf {\bibinfo
  {volume} {98}},\ \bibinfo {pages} {267802} (\bibinfo {year}
  {2007})}\BibitemShut {NoStop}%
\bibitem [{\citenamefont {Rockford}\ \emph {et~al.}(1999)\citenamefont
  {Rockford}, \citenamefont {Liu}, \citenamefont {Mansky}, \citenamefont
  {Russell}, \citenamefont {Yoon},\ and\ \citenamefont
  {Mochrie}}]{Rockford1999}%
  \BibitemOpen
  \bibfield  {author} {\bibinfo {author} {\bibfnamefont {L.}~\bibnamefont
  {Rockford}}, \bibinfo {author} {\bibfnamefont {Y.}~\bibnamefont {Liu}},
  \bibinfo {author} {\bibfnamefont {P.}~\bibnamefont {Mansky}}, \bibinfo
  {author} {\bibfnamefont {T.}~\bibnamefont {Russell}}, \bibinfo {author}
  {\bibfnamefont {M.}~\bibnamefont {Yoon}}, \ and\ \bibinfo {author}
  {\bibfnamefont {S.}~\bibnamefont {Mochrie}},\ }\href@noop {} {\bibfield
  {journal} {\bibinfo  {journal} {Phys. Rev. Lett.}\ }\textbf {\bibinfo
  {volume} {82}},\ \bibinfo {pages} {2602} (\bibinfo {year}
  {1999})}\BibitemShut {NoStop}%
\bibitem [{\citenamefont {Higgins}\ and\ \citenamefont
  {Jones}(2000)}]{Higgins2000}%
  \BibitemOpen
  \bibfield  {author} {\bibinfo {author} {\bibfnamefont {A.~M.}\ \bibnamefont
  {Higgins}}\ and\ \bibinfo {author} {\bibfnamefont {R.~A.}\ \bibnamefont
  {Jones}},\ }\href@noop {} {\bibfield  {journal} {\bibinfo  {journal}
  {Nature}\ }\textbf {\bibinfo {volume} {404}},\ \bibinfo {pages} {476}
  (\bibinfo {year} {2000})}\BibitemShut {NoStop}%
\bibitem [{\citenamefont {Konnur}\ \emph {et~al.}(2000)\citenamefont {Konnur},
  \citenamefont {Kargupta},\ and\ \citenamefont {Sharma}}]{Konnur2000}%
  \BibitemOpen
  \bibfield  {author} {\bibinfo {author} {\bibfnamefont {R.}~\bibnamefont
  {Konnur}}, \bibinfo {author} {\bibfnamefont {K.}~\bibnamefont {Kargupta}}, \
  and\ \bibinfo {author} {\bibfnamefont {A.}~\bibnamefont {Sharma}},\
  }\href@noop {} {\bibfield  {journal} {\bibinfo  {journal} {Phys. Rev. Lett.}\
  }\textbf {\bibinfo {volume} {84}},\ \bibinfo {pages} {931} (\bibinfo {year}
  {2000})}\BibitemShut {NoStop}%
\bibitem [{\citenamefont {Kargupta}\ and\ \citenamefont
  {Sharma}(2002)}]{Kargupta2002}%
  \BibitemOpen
  \bibfield  {author} {\bibinfo {author} {\bibfnamefont {K.}~\bibnamefont
  {Kargupta}}\ and\ \bibinfo {author} {\bibfnamefont {A.}~\bibnamefont
  {Sharma}},\ }\href@noop {} {\bibfield  {journal} {\bibinfo  {journal}
  {Langmuir}\ }\textbf {\bibinfo {volume} {18}},\ \bibinfo {pages} {1893}
  (\bibinfo {year} {2002})}\BibitemShut {NoStop}%
\bibitem [{\citenamefont {Suh}\ and\ \citenamefont {Lee}(2002)}]{Suh2002}%
  \BibitemOpen
  \bibfield  {author} {\bibinfo {author} {\bibfnamefont {K.~Y.}\ \bibnamefont
  {Suh}}\ and\ \bibinfo {author} {\bibfnamefont {H.~H.}\ \bibnamefont {Lee}},\
  }\href@noop {} {\bibfield  {journal} {\bibinfo  {journal} {Adv. Funct.
  Mater.}\ }\textbf {\bibinfo {volume} {12}},\ \bibinfo {pages} {405} (\bibinfo
  {year} {2002})}\BibitemShut {NoStop}%
\bibitem [{\citenamefont {Sehgal}\ \emph {et~al.}(2002)\citenamefont {Sehgal},
  \citenamefont {Ferreiro}, \citenamefont {Douglas}, \citenamefont {Amis},\
  and\ \citenamefont {Karim}}]{Sehgal2002}%
  \BibitemOpen
  \bibfield  {author} {\bibinfo {author} {\bibfnamefont {A.}~\bibnamefont
  {Sehgal}}, \bibinfo {author} {\bibfnamefont {V.}~\bibnamefont {Ferreiro}},
  \bibinfo {author} {\bibfnamefont {J.~F.}\ \bibnamefont {Douglas}}, \bibinfo
  {author} {\bibfnamefont {E.~J.}\ \bibnamefont {Amis}}, \ and\ \bibinfo
  {author} {\bibfnamefont {A.}~\bibnamefont {Karim}},\ }\href@noop {}
  {\bibfield  {journal} {\bibinfo  {journal} {Langmuir}\ }\textbf {\bibinfo
  {volume} {18}},\ \bibinfo {pages} {7041} (\bibinfo {year}
  {2002})}\BibitemShut {NoStop}%
\bibitem [{\citenamefont {Shanahan}(1987)}]{Shanahan1987}%
  \BibitemOpen
  \bibfield  {author} {\bibinfo {author} {\bibfnamefont {M.}~\bibnamefont
  {Shanahan}},\ }\href@noop {} {\bibfield  {journal} {\bibinfo  {journal} {J.
  Phys. D}\ }\textbf {\bibinfo {volume} {20}},\ \bibinfo {pages} {945}
  (\bibinfo {year} {1987})}\BibitemShut {NoStop}%
\bibitem [{\citenamefont {Marchand}\ \emph {et~al.}(2012)\citenamefont
  {Marchand}, \citenamefont {Das}, \citenamefont {Snoeijer},\ and\
  \citenamefont {Andreotti}}]{Marchand2012a}%
  \BibitemOpen
  \bibfield  {author} {\bibinfo {author} {\bibfnamefont {A.}~\bibnamefont
  {Marchand}}, \bibinfo {author} {\bibfnamefont {S.}~\bibnamefont {Das}},
  \bibinfo {author} {\bibfnamefont {J.~H.}\ \bibnamefont {Snoeijer}}, \ and\
  \bibinfo {author} {\bibfnamefont {B.}~\bibnamefont {Andreotti}},\ }\href
  {\doibase 10.1103/PhysRevLett.109.236101} {\bibfield  {journal} {\bibinfo
  {journal} {Phys. Rev. Lett.}\ }\textbf {\bibinfo {volume} {109}},\ \bibinfo
  {pages} {236101} (\bibinfo {year} {2012})}\BibitemShut {NoStop}%
\bibitem [{\citenamefont {Style}\ \emph
  {et~al.}(2013{\natexlab{a}})\citenamefont {Style}, \citenamefont {Che},
  \citenamefont {Park}, \citenamefont {Weon}, \citenamefont {Je}, \citenamefont
  {Hyland}, \citenamefont {German}, \citenamefont {Power}, \citenamefont
  {Wilen}, \citenamefont {Wettlaufer},\ and\ \citenamefont
  {Dufresne}}]{Style2013}%
  \BibitemOpen
  \bibfield  {author} {\bibinfo {author} {\bibfnamefont {R.~W.}\ \bibnamefont
  {Style}}, \bibinfo {author} {\bibfnamefont {Y.}~\bibnamefont {Che}}, \bibinfo
  {author} {\bibfnamefont {S.~J.}\ \bibnamefont {Park}}, \bibinfo {author}
  {\bibfnamefont {B.~M.}\ \bibnamefont {Weon}}, \bibinfo {author}
  {\bibfnamefont {J.~H.}\ \bibnamefont {Je}}, \bibinfo {author} {\bibfnamefont
  {C.}~\bibnamefont {Hyland}}, \bibinfo {author} {\bibfnamefont {G.~K.}\
  \bibnamefont {German}}, \bibinfo {author} {\bibfnamefont {M.~P.}\
  \bibnamefont {Power}}, \bibinfo {author} {\bibfnamefont {L.~A.}\ \bibnamefont
  {Wilen}}, \bibinfo {author} {\bibfnamefont {J.~S.}\ \bibnamefont
  {Wettlaufer}}, \ and\ \bibinfo {author} {\bibfnamefont {E.}~\bibnamefont
  {Dufresne}},\ }\href@noop {} {\bibfield  {journal} {\bibinfo  {journal}
  {Proc. Natl. Acad. Sci. U.S.A.}\ }\textbf {\bibinfo {volume} {110}},\
  \bibinfo {pages} {12541} (\bibinfo {year} {2013}{\natexlab{a}})}\BibitemShut
  {NoStop}%
\bibitem [{\citenamefont {Style}\ \emph
  {et~al.}(2013{\natexlab{b}})\citenamefont {Style}, \citenamefont
  {Boltyanskiy}, \citenamefont {Che}, \citenamefont {Wettlaufer}, \citenamefont
  {Wilen},\ and\ \citenamefont {Dufresne}}]{Style2013a}%
  \BibitemOpen
  \bibfield  {author} {\bibinfo {author} {\bibfnamefont {R.}~\bibnamefont
  {Style}}, \bibinfo {author} {\bibfnamefont {R.}~\bibnamefont {Boltyanskiy}},
  \bibinfo {author} {\bibfnamefont {Y.}~\bibnamefont {Che}}, \bibinfo {author}
  {\bibfnamefont {J.}~\bibnamefont {Wettlaufer}}, \bibinfo {author}
  {\bibfnamefont {L.~A.}\ \bibnamefont {Wilen}}, \ and\ \bibinfo {author}
  {\bibfnamefont {E.}~\bibnamefont {Dufresne}},\ }\href {\doibase
  10.1103/PhysRevLett.110.066103} {\bibfield  {journal} {\bibinfo  {journal}
  {Phys. Rev. Lett.}\ }\textbf {\bibinfo {volume} {110}},\ \bibinfo {pages}
  {066103} (\bibinfo {year} {2013}{\natexlab{b}})}\BibitemShut {NoStop}%
\bibitem [{\citenamefont {Nadermann}\ \emph {et~al.}(2013)\citenamefont
  {Nadermann}, \citenamefont {Hui},\ and\ \citenamefont
  {Jagota}}]{Nadermann2013}%
  \BibitemOpen
  \bibfield  {author} {\bibinfo {author} {\bibfnamefont {N.}~\bibnamefont
  {Nadermann}}, \bibinfo {author} {\bibfnamefont {C.-Y.}\ \bibnamefont {Hui}},
  \ and\ \bibinfo {author} {\bibfnamefont {A.}~\bibnamefont {Jagota}},\ }\href
  {\doibase 10.1073/pnas.1304587110} {\bibfield  {journal} {\bibinfo  {journal}
  {Proc. Natl. Acad. Sci. U.S.A.}\ }\textbf {\bibinfo {volume} {110}},\
  \bibinfo {pages} {10541} (\bibinfo {year} {2013})}\BibitemShut {NoStop}%
\bibitem [{\citenamefont {Park}\ \emph {et~al.}(2014)\citenamefont {Park},
  \citenamefont {Weon}, \citenamefont {Lee}, \citenamefont {Lee}, \citenamefont
  {Kim},\ and\ \citenamefont {Je}}]{Park2014}%
  \BibitemOpen
  \bibfield  {author} {\bibinfo {author} {\bibfnamefont {S.~J.}\ \bibnamefont
  {Park}}, \bibinfo {author} {\bibfnamefont {B.~M.}\ \bibnamefont {Weon}},
  \bibinfo {author} {\bibfnamefont {J.~S.}\ \bibnamefont {Lee}}, \bibinfo
  {author} {\bibfnamefont {J.}~\bibnamefont {Lee}}, \bibinfo {author}
  {\bibfnamefont {J.}~\bibnamefont {Kim}}, \ and\ \bibinfo {author}
  {\bibfnamefont {J.~H.}\ \bibnamefont {Je}},\ }\href {\doibase
  10.1038/ncomms5369} {\bibfield  {journal} {\bibinfo  {journal} {Nat. Comm.}\
  }\textbf {\bibinfo {volume} {5}},\ \bibinfo {pages} {4369} (\bibinfo {year}
  {2014})}\BibitemShut {NoStop}%
\bibitem [{\citenamefont {Bostwick}\ \emph {et~al.}(2014)\citenamefont
  {Bostwick}, \citenamefont {Shearer},\ and\ \citenamefont
  {Daniels}}]{Bostwick2014}%
  \BibitemOpen
  \bibfield  {author} {\bibinfo {author} {\bibfnamefont {J.~B.}\ \bibnamefont
  {Bostwick}}, \bibinfo {author} {\bibfnamefont {M.}~\bibnamefont {Shearer}}, \
  and\ \bibinfo {author} {\bibfnamefont {K.~E.}\ \bibnamefont {Daniels}},\
  }\href@noop {} {\bibfield  {journal} {\bibinfo  {journal} {Soft Matter}\
  }\textbf {\bibinfo {volume} {10}},\ \bibinfo {pages} {7361} (\bibinfo {year}
  {2014})}\BibitemShut {NoStop}%
\bibitem [{\citenamefont {Schulman}\ and\ \citenamefont
  {Dalnoki-Veress}(2015)}]{Schulman2016}%
  \BibitemOpen
  \bibfield  {author} {\bibinfo {author} {\bibfnamefont {R.~D.}\ \bibnamefont
  {Schulman}}\ and\ \bibinfo {author} {\bibfnamefont {K.}~\bibnamefont
  {Dalnoki-Veress}},\ }\href {\doibase 10.1103/PhysRevLett.115.206101}
  {\bibfield  {journal} {\bibinfo  {journal} {Phys. Rev. Lett.}\ }\textbf
  {\bibinfo {volume} {115}},\ \bibinfo {pages} {206101} (\bibinfo {year}
  {2015})}\BibitemShut {NoStop}%
\bibitem [{\citenamefont {Schulman}\ \emph {et~al.}(2017)\citenamefont
  {Schulman}, \citenamefont {Ledesma-Alonso}, \citenamefont {Salez},
  \citenamefont {Rapha\"el},\ and\ \citenamefont
  {Dalnoki-Veress}}]{Schulman2017}%
  \BibitemOpen
  \bibfield  {author} {\bibinfo {author} {\bibfnamefont {R.~D.}\ \bibnamefont
  {Schulman}}, \bibinfo {author} {\bibfnamefont {R.}~\bibnamefont
  {Ledesma-Alonso}}, \bibinfo {author} {\bibfnamefont {T.}~\bibnamefont
  {Salez}}, \bibinfo {author} {\bibfnamefont {E.}~\bibnamefont {Rapha\"el}}, \
  and\ \bibinfo {author} {\bibfnamefont {K.}~\bibnamefont {Dalnoki-Veress}},\
  }\href {\doibase 10.1103/PhysRevLett.118.198002} {\bibfield  {journal}
  {\bibinfo  {journal} {Phys. Rev. Lett.}\ }\textbf {\bibinfo {volume} {118}},\
  \bibinfo {pages} {198002} (\bibinfo {year} {2017})}\BibitemShut {NoStop}%
\bibitem [{\citenamefont {Martin}\ and\ \citenamefont
  {Brochard-Wyart}(1998)}]{Martin1998}%
  \BibitemOpen
  \bibfield  {author} {\bibinfo {author} {\bibfnamefont {P.}~\bibnamefont
  {Martin}}\ and\ \bibinfo {author} {\bibfnamefont {F.}~\bibnamefont
  {Brochard-Wyart}},\ }\href@noop {} {\bibfield  {journal} {\bibinfo  {journal}
  {Phys. Rev. Lett.}\ }\textbf {\bibinfo {volume} {80}},\ \bibinfo {pages}
  {3296} (\bibinfo {year} {1998})}\BibitemShut {NoStop}%
\bibitem [{\citenamefont {Martin}\ \emph {et~al.}(2000)\citenamefont {Martin},
  \citenamefont {Rossier}, \citenamefont {Buguin}, \citenamefont {Auroy},\ and\
  \citenamefont {Brochard-Wyart}}]{Martin2000}%
  \BibitemOpen
  \bibfield  {author} {\bibinfo {author} {\bibfnamefont {A.}~\bibnamefont
  {Martin}}, \bibinfo {author} {\bibfnamefont {O.}~\bibnamefont {Rossier}},
  \bibinfo {author} {\bibfnamefont {A.}~\bibnamefont {Buguin}}, \bibinfo
  {author} {\bibfnamefont {P.}~\bibnamefont {Auroy}}, \ and\ \bibinfo {author}
  {\bibfnamefont {F.}~\bibnamefont {Brochard-Wyart}},\ }\href@noop {}
  {\bibfield  {journal} {\bibinfo  {journal} {Eur. Phys. J. E}\ }\textbf
  {\bibinfo {volume} {3}},\ \bibinfo {pages} {337} (\bibinfo {year}
  {2000})}\BibitemShut {NoStop}%
\bibitem [{\citenamefont {Martin}\ \emph {et~al.}(2001)\citenamefont {Martin},
  \citenamefont {Buguin},\ and\ \citenamefont {Brochard-Wyart}}]{Martin2001}%
  \BibitemOpen
  \bibfield  {author} {\bibinfo {author} {\bibfnamefont {A.}~\bibnamefont
  {Martin}}, \bibinfo {author} {\bibfnamefont {A.}~\bibnamefont {Buguin}}, \
  and\ \bibinfo {author} {\bibfnamefont {F.}~\bibnamefont {Brochard-Wyart}},\
  }\href@noop {} {\bibfield  {journal} {\bibinfo  {journal} {Langmuir}\
  }\textbf {\bibinfo {volume} {17}},\ \bibinfo {pages} {6553} (\bibinfo {year}
  {2001})}\BibitemShut {NoStop}%
\bibitem [{\citenamefont {G{\'e}rardin}\ \emph {et~al.}(2005)\citenamefont
  {G{\'e}rardin}, \citenamefont {Verneuil}, \citenamefont {Constant},
  \citenamefont {Dubois}, \citenamefont {Clain}, \citenamefont {Noblin},
  \citenamefont {Buguin},\ and\ \citenamefont {Brochard-Wyart}}]{Gerardin2005}%
  \BibitemOpen
  \bibfield  {author} {\bibinfo {author} {\bibfnamefont {H.}~\bibnamefont
  {G{\'e}rardin}}, \bibinfo {author} {\bibfnamefont {E.}~\bibnamefont
  {Verneuil}}, \bibinfo {author} {\bibfnamefont {A.}~\bibnamefont {Constant}},
  \bibinfo {author} {\bibfnamefont {S.}~\bibnamefont {Dubois}}, \bibinfo
  {author} {\bibfnamefont {J.}~\bibnamefont {Clain}}, \bibinfo {author}
  {\bibfnamefont {X.}~\bibnamefont {Noblin}}, \bibinfo {author} {\bibfnamefont
  {A.}~\bibnamefont {Buguin}}, \ and\ \bibinfo {author} {\bibfnamefont
  {F.}~\bibnamefont {Brochard-Wyart}},\ }\href@noop {} {\bibfield  {journal}
  {\bibinfo  {journal} {Eur. Phys. Lett.}\ }\textbf {\bibinfo {volume} {71}},\
  \bibinfo {pages} {418} (\bibinfo {year} {2005})}\BibitemShut {NoStop}%
\bibitem [{\citenamefont {Qiao}\ and\ \citenamefont {He}(2008)}]{Qiao2008}%
  \BibitemOpen
  \bibfield  {author} {\bibinfo {author} {\bibfnamefont {L.}~\bibnamefont
  {Qiao}}\ and\ \bibinfo {author} {\bibfnamefont {L.}~\bibnamefont {He}},\
  }\href@noop {} {\bibfield  {journal} {\bibinfo  {journal} {Eur. Phys. J. E}\
  }\textbf {\bibinfo {volume} {26}},\ \bibinfo {pages} {387} (\bibinfo {year}
  {2008})}\BibitemShut {NoStop}%
\bibitem [{\citenamefont {Becker}\ and\ \citenamefont
  {Mugele}(2002)}]{Becker2002}%
  \BibitemOpen
  \bibfield  {author} {\bibinfo {author} {\bibfnamefont {T.}~\bibnamefont
  {Becker}}\ and\ \bibinfo {author} {\bibfnamefont {F.}~\bibnamefont
  {Mugele}},\ }\href@noop {} {\bibfield  {journal} {\bibinfo  {journal} {J.
  Phys. Condens. Matter}\ }\textbf {\bibinfo {volume} {15}},\ \bibinfo {pages}
  {S321} (\bibinfo {year} {2002})}\BibitemShut {NoStop}%
\bibitem [{\citenamefont {Kumar}\ and\ \citenamefont
  {Matar}(2004)}]{Kumar2004}%
  \BibitemOpen
  \bibfield  {author} {\bibinfo {author} {\bibfnamefont {S.}~\bibnamefont
  {Kumar}}\ and\ \bibinfo {author} {\bibfnamefont {O.~K.}\ \bibnamefont
  {Matar}},\ }\href@noop {} {\bibfield  {journal} {\bibinfo  {journal} {J.
  Colloid Interface Sci.}\ }\textbf {\bibinfo {volume} {273}},\ \bibinfo
  {pages} {581} (\bibinfo {year} {2004})}\BibitemShut {NoStop}%
\bibitem [{\citenamefont {de~Beer}\ \emph {et~al.}(2009)\citenamefont
  {de~Beer}, \citenamefont {{D. `t Mannetje}}, \citenamefont {Zantema},\ and\
  \citenamefont {Mugele}}]{deBeer2009}%
  \BibitemOpen
  \bibfield  {author} {\bibinfo {author} {\bibfnamefont {S.}~\bibnamefont
  {de~Beer}}, \bibinfo {author} {\bibnamefont {{D. `t Mannetje}}}, \bibinfo
  {author} {\bibfnamefont {S.}~\bibnamefont {Zantema}}, \ and\ \bibinfo
  {author} {\bibfnamefont {F.}~\bibnamefont {Mugele}},\ }\href@noop {}
  {\bibfield  {journal} {\bibinfo  {journal} {Langmuir}\ }\textbf {\bibinfo
  {volume} {26}},\ \bibinfo {pages} {3280} (\bibinfo {year}
  {2009})}\BibitemShut {NoStop}%
\bibitem [{\citenamefont {Carlson}\ \emph {et~al.}(2015)\citenamefont
  {Carlson}, \citenamefont {Mandre},\ and\ \citenamefont
  {Mahadevan}}]{Carlson2015}%
  \BibitemOpen
  \bibfield  {author} {\bibinfo {author} {\bibfnamefont {A.}~\bibnamefont
  {Carlson}}, \bibinfo {author} {\bibfnamefont {S.}~\bibnamefont {Mandre}}, \
  and\ \bibinfo {author} {\bibfnamefont {L.}~\bibnamefont {Mahadevan}},\
  }\href@noop {} {\bibfield  {journal} {\bibinfo  {journal} {arXiv}\ }
  (\bibinfo {year} {2015})}\BibitemShut {NoStop}%
\bibitem [{SI()}]{SI}%
  \BibitemOpen
  \href@noop {} {}\bibinfo {note} {See supplemental material for a more detailed overview of the sample
  preparation, equations to estimate tension and film thickness upon straining,
  late stage evolution of elongated holes, further explanation of the force
  balance at the rim's contact line, and movies for the growth of circular and
  square holes.}\BibitemShut {Stop}%
\bibitem [{\citenamefont {Jacobs}\ \emph {et~al.}(1998)\citenamefont {Jacobs},
  \citenamefont {Seemann}, \citenamefont {Schatz},\ and\ \citenamefont
  {Herminghaus}}]{Jacobs1998}%
  \BibitemOpen
  \bibfield  {author} {\bibinfo {author} {\bibfnamefont {K.}~\bibnamefont
  {Jacobs}}, \bibinfo {author} {\bibfnamefont {R.}~\bibnamefont {Seemann}},
  \bibinfo {author} {\bibfnamefont {G.}~\bibnamefont {Schatz}}, \ and\ \bibinfo
  {author} {\bibfnamefont {S.}~\bibnamefont {Herminghaus}},\ }\href {\doibase
  10.1021/la9804435} {\bibfield  {journal} {\bibinfo  {journal} {Langmuir}\
  }\textbf {\bibinfo {volume} {14}},\ \bibinfo {pages} {4961} (\bibinfo {year}
  {1998})}\BibitemShut {NoStop}%
\bibitem [{\citenamefont {Ghatak}\ \emph {et~al.}(1999)\citenamefont {Ghatak},
  \citenamefont {Khanna},\ and\ \citenamefont {Sharma}}]{Ghatak1999}%
  \BibitemOpen
  \bibfield  {author} {\bibinfo {author} {\bibfnamefont {A.}~\bibnamefont
  {Ghatak}}, \bibinfo {author} {\bibfnamefont {R.}~\bibnamefont {Khanna}}, \
  and\ \bibinfo {author} {\bibfnamefont {A.}~\bibnamefont {Sharma}},\ }\href
  {\doibase http://dx.doi.org/10.1006/jcis.1998.6052} {\bibfield  {journal}
  {\bibinfo  {journal} {J. Colloid Interface Sci.}\ }\textbf {\bibinfo {volume}
  {212}},\ \bibinfo {pages} {483 } (\bibinfo {year} {1999})}\BibitemShut
  {NoStop}%
\bibitem [{\citenamefont {M{\"u}nch}(2005)}]{Munch2005}%
  \BibitemOpen
  \bibfield  {author} {\bibinfo {author} {\bibfnamefont {A.}~\bibnamefont
  {M{\"u}nch}},\ }\href@noop {} {\bibfield  {journal} {\bibinfo  {journal} {J.
  Phys. Cond. Mat.}\ }\textbf {\bibinfo {volume} {17}},\ \bibinfo {pages}
  {S309} (\bibinfo {year} {2005})}\BibitemShut {NoStop}%
\bibitem [{\citenamefont {Sharma}\ and\ \citenamefont
  {Reiter}(1996)}]{Sharma1996}%
  \BibitemOpen
  \bibfield  {author} {\bibinfo {author} {\bibfnamefont {A.}~\bibnamefont
  {Sharma}}\ and\ \bibinfo {author} {\bibfnamefont {G.}~\bibnamefont
  {Reiter}},\ }\href@noop {} {\bibfield  {journal} {\bibinfo  {journal} {J.
  Colloid Interface Sci.}\ }\textbf {\bibinfo {volume} {178}},\ \bibinfo
  {pages} {383} (\bibinfo {year} {1996})}\BibitemShut {NoStop}%
\bibitem [{\citenamefont {Choi}\ and\ \citenamefont
  {Zhang~Newby}(2006)}]{Choi2006}%
  \BibitemOpen
  \bibfield  {author} {\bibinfo {author} {\bibfnamefont {S.-H.}\ \bibnamefont
  {Choi}}\ and\ \bibinfo {author} {\bibfnamefont {B.-m.}\ \bibnamefont
  {Zhang~Newby}},\ }\href@noop {} {\bibfield  {journal} {\bibinfo  {journal}
  {J. Chem. Phys.}\ }\textbf {\bibinfo {volume} {124}},\ \bibinfo {pages}
  {054702} (\bibinfo {year} {2006})}\BibitemShut {NoStop}%
\bibitem [{\citenamefont {B{\"a}umchen}\ \emph {et~al.}(2014)\citenamefont
  {B{\"a}umchen}, \citenamefont {Marquant}, \citenamefont {Blossey},
  \citenamefont {M{\"u}nch}, \citenamefont {Wagner},\ and\ \citenamefont
  {Jacobs}}]{Baumchen2014}%
  \BibitemOpen
  \bibfield  {author} {\bibinfo {author} {\bibfnamefont {O.}~\bibnamefont
  {B{\"a}umchen}}, \bibinfo {author} {\bibfnamefont {L.}~\bibnamefont
  {Marquant}}, \bibinfo {author} {\bibfnamefont {R.}~\bibnamefont {Blossey}},
  \bibinfo {author} {\bibfnamefont {A.}~\bibnamefont {M{\"u}nch}}, \bibinfo
  {author} {\bibfnamefont {B.}~\bibnamefont {Wagner}}, \ and\ \bibinfo {author}
  {\bibfnamefont {K.}~\bibnamefont {Jacobs}},\ }\href@noop {} {\bibfield
  {journal} {\bibinfo  {journal} {Phys. Rev. Lett.}\ }\textbf {\bibinfo
  {volume} {113}},\ \bibinfo {pages} {014501} (\bibinfo {year}
  {2014})}\BibitemShut {NoStop}%
\bibitem [{\citenamefont {Mi}\ \emph {et~al.}(2014)\citenamefont {Mi},
  \citenamefont {Jing}, \citenamefont {Salick}, \citenamefont {Crone},
  \citenamefont {Peng},\ and\ \citenamefont {Turng}}]{elastollan_modulus3}%
  \BibitemOpen
  \bibfield  {author} {\bibinfo {author} {\bibfnamefont {H.-Y.}\ \bibnamefont
  {Mi}}, \bibinfo {author} {\bibfnamefont {X.}~\bibnamefont {Jing}}, \bibinfo
  {author} {\bibfnamefont {M.~R.}\ \bibnamefont {Salick}}, \bibinfo {author}
  {\bibfnamefont {W.~C.}\ \bibnamefont {Crone}}, \bibinfo {author}
  {\bibfnamefont {X.-F.}\ \bibnamefont {Peng}}, \ and\ \bibinfo {author}
  {\bibfnamefont {L.-S.}\ \bibnamefont {Turng}},\ }\href@noop {} {\bibfield
  {journal} {\bibinfo  {journal} {Adv. Polym. Tech.}\ }\textbf {\bibinfo
  {volume} {33}} (\bibinfo {year} {2014})}\BibitemShut {NoStop}%
\end{thebibliography}

\begin{thebibliography}{2}%
\makeatletter
\providecommand \@ifxundefined [1]{%
 \@ifx{#1\undefined}
}%
\providecommand \@ifnum [1]{%
 \ifnum #1\expandafter \@firstoftwo
 \else \expandafter \@secondoftwo
 \fi
}%
\providecommand \@ifx [1]{%
 \ifx #1\expandafter \@firstoftwo
 \else \expandafter \@secondoftwo
 \fi
}%
\providecommand \natexlab [1]{#1}%
\providecommand \enquote  [1]{``#1''}%
\providecommand \bibnamefont  [1]{#1}%
\providecommand \bibfnamefont [1]{#1}%
\providecommand \citenamefont [1]{#1}%
\providecommand \href@noop [0]{\@secondoftwo}%
\providecommand \href [0]{\begingroup \@sanitize@url \@href}%
\providecommand \@href[1]{\@@startlink{#1}\@@href}%
\providecommand \@@href[1]{\endgroup#1\@@endlink}%
\providecommand \@sanitize@url [0]{\catcode `\\12\catcode `\$12\catcode
  `\&12\catcode `\#12\catcode `\^12\catcode `\_12\catcode `\%12\relax}%
\providecommand \@@startlink[1]{}%
\providecommand \@@endlink[0]{}%
\providecommand \url  [0]{\begingroup\@sanitize@url \@url }%
\providecommand \@url [1]{\endgroup\@href {#1}{\urlprefix }}%
\providecommand \urlprefix  [0]{URL }%
\providecommand \Eprint [0]{\href }%
\providecommand \doibase [0]{http://dx.doi.org/}%
\providecommand \selectlanguage [0]{\@gobble}%
\providecommand \bibinfo  [0]{\@secondoftwo}%
\providecommand \bibfield  [0]{\@secondoftwo}%
\providecommand \translation [1]{[#1]}%
\providecommand \BibitemOpen [0]{}%
\providecommand \bibitemStop [0]{}%
\providecommand \bibitemNoStop [0]{.\EOS\space}%
\providecommand \EOS [0]{\spacefactor3000\relax}%
\providecommand \BibitemShut  [1]{\csname bibitem#1\endcsname}%
\let\auto@bib@innerbib\@empty
\bibitem [{\citenamefont {Timoshenko}\ and\ \citenamefont
  {Goodier}(1951)}]{timoshenko1951}%
  \BibitemOpen
  \bibfield  {author} {\bibinfo {author} {\bibfnamefont {S.}~\bibnamefont
  {Timoshenko}}\ and\ \bibinfo {author} {\bibfnamefont {J.}~\bibnamefont
  {Goodier}},\ }\href@noop {} {\emph {\bibinfo {title} {Theory of Elasticity,
  2nd edition}}}\ (\bibinfo  {publisher} {McGraw-Hill Book Company, Inc., New
  York},\ \bibinfo {year} {1951})\BibitemShut {NoStop}%
\bibitem [{\citenamefont {Mi}\ \emph {et~al.}(2014)\citenamefont {Mi},
  \citenamefont {Jing}, \citenamefont {Salick}, \citenamefont {Crone},
  \citenamefont {Peng},\ and\ \citenamefont {Turng}}]{elastollan_modulus3}%
  \BibitemOpen
  \bibfield  {author} {\bibinfo {author} {\bibfnamefont {H.-Y.}\ \bibnamefont
  {Mi}}, \bibinfo {author} {\bibfnamefont {X.}~\bibnamefont {Jing}}, \bibinfo
  {author} {\bibfnamefont {M.~R.}\ \bibnamefont {Salick}}, \bibinfo {author}
  {\bibfnamefont {W.~C.}\ \bibnamefont {Crone}}, \bibinfo {author}
  {\bibfnamefont {X.-F.}\ \bibnamefont {Peng}}, \ and\ \bibinfo {author}
  {\bibfnamefont {L.-S.}\ \bibnamefont {Turng}},\ }\href@noop {} {\bibfield
  {journal} {\bibinfo  {journal} {Adv. Polym. Tech.}\ }\textbf {\bibinfo
  {volume} {33}} (\bibinfo {year} {2014})}\BibitemShut {NoStop}%
\end{thebibliography}
%

\pagebreak
\widetext

\renewcommand{\thefigure}{S\arabic{figure}}
\setcounter{figure}{0}
\renewcommand{\theequation}{S\arabic{equation}}
\setcounter{equation}{0}

\section{Supplemental Information for: ``Liquid dewetting under a thin elastic film"} 


\section{Sample Preparation}
\begin{figure}[b]
     \includegraphics[width=17cm]{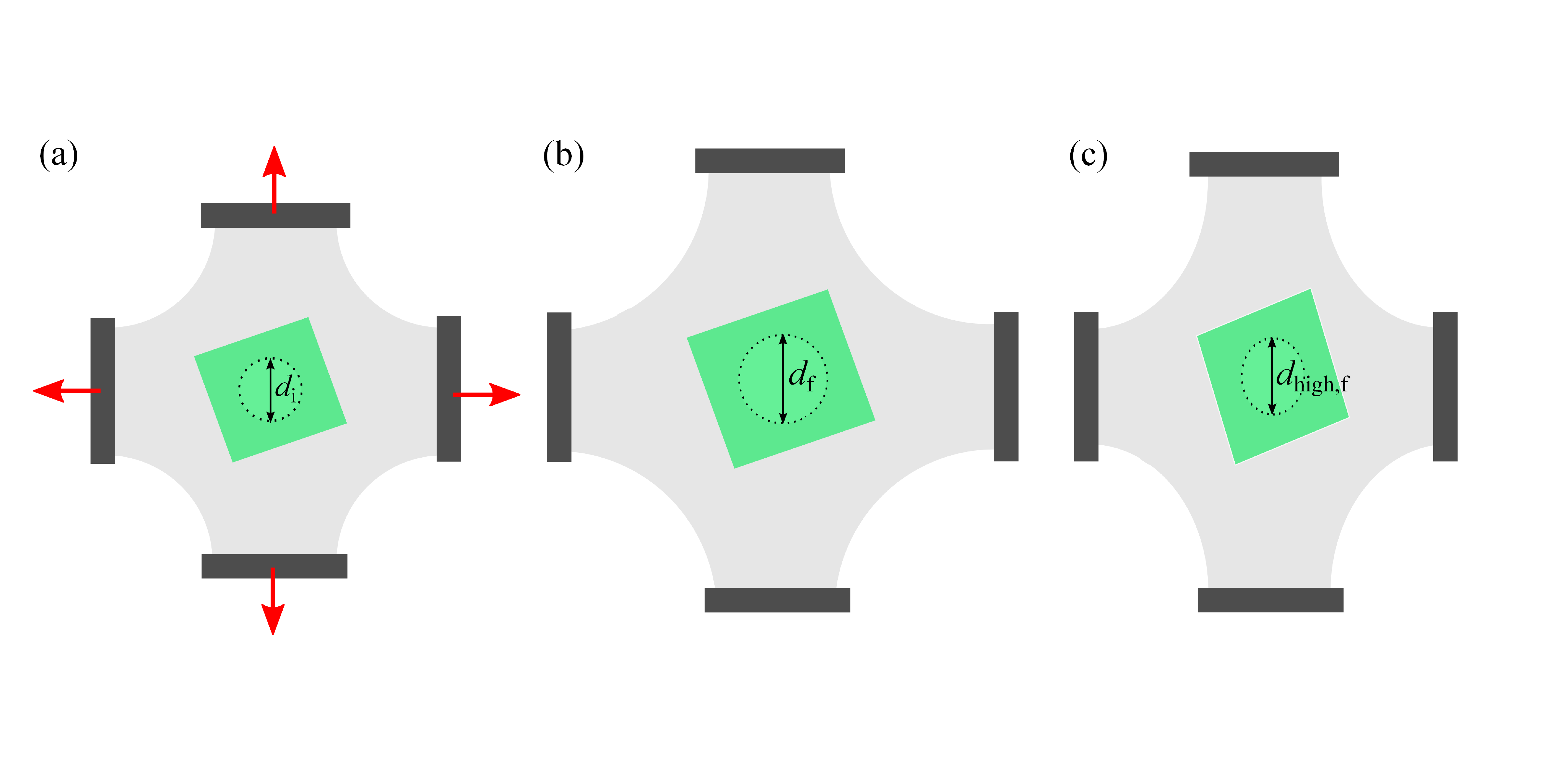}
\caption{(a) Top view schematic of the straining apparatus. The Elastollan film is free-standing over the hole in the Elastosil with diameter $d_\mathrm{i}$. The clamped sides of the Elastosil can be translated along the axes indicated by the arrows. (b) An Elastollan film which has been strained equally in both directions (isotropic tension) to a final diameter of $d_\mathrm{f}$. (c) An Elastollan film which has been strained biaxially to a final diameter $d_\mathrm{high,f}$ in one direction, while being held fixed in the orthogonal direction.}
\label{figS1}
\end{figure}

In these experiments, thin polystyrene (PS) films (Scientific Polymer Products, number averaged molecular weight $M_n$ = 15,800 g/mol, polydispersity index 1.05) of thickness $h_\mathrm{PS} \sim 100$~nm, measured using ellipsometry (Accurion, EP3), were prepared through spincoating out of a toluene solution onto 1~cm x 1~cm silicon wafers. The PS was removed near the four sides of the silicon using an acetone-wetted cotton swab. The PS samples were then annealed on a hot stage (Linkam) at 140$^\circ$C for 10 min to relax the polymer chains and remove any residual solvent.  Elastomeric films were prepared from Elastollan TPU 1185A (BASF). Upon spincoating these solutions, the Elastollan polymers, which contain hard and soft segments, self-assemble to form an elastomer with physical crosslinks. The Elastollan/cyclohexanone solutions were cast onto 3~cm x 3~cm freshly cleaved mica substrates (Ted Pella Inc.) to produce highly uniform ($<$5\% variation) films with initial thickness in the range $h_0 \sim $  50 -- 320 nm, measured using ellipsometry. These films were subsequently heated atop a hot stage at 150$^\circ$C for 10 min to remove any residual solvent from the elastomer. After annealing, these films were scored using a scalpel along each edge of the mica and  dipped into an ultrapure water bath (18.2 M$\Omega \cdot$cm, Pall, Cascada, LS). In doing so, a thin film of water wedges itself between the Elastollan film and the mica substrate. This sample is then transferred to a home made straining apparatus, depicted in Fig.~\ref{figS1}(a). The apparatus consists of a 250~$\mu$m thick Elastosil (Wacker Chemie) sheet which has been cut into a rounded plus sign shape, but contains a circular hole at its center. The Elastosil film is clamped at its four sides and supported by four posts which are able to translate in the directions indicated by arrows in Fig.~\ref{figS1}(a). The Elastollan sample is placed atop of the hole in the Elastosil, which immediately causes the the Elastollan film to form strong contact with the Elastosil, allowing the mica to be peeled off and removed. Thus, a thin Elastollan film is left free-standing over the hole in the Elastosil. 

Upon transfer, the films have a small ($<$~5\%) pre-strain due to being stretched by the surface tension of the water during the sample preparation. In order for the initial condition to be an unstrained film, we must relieve this pre-strain. This is done by slowly bringing all four supports inwards to shrink the size of the Elastosil hole. At the point that the pre-strain is relieved, wrinkles just begin to appear in the Elastollan film. This point serves as the initial condition for the Elastollan film. To then strain the Elastollan film, the four posts supporting the Elastosil are moved outwards to stretch the Elastosil film, and hence, expand the hole at its center. To generate isotropic tension, each support is moved an equal distance such that the Elastosil hole remains circular in shape (Fig.~\ref{figS1}(b)). For the biaxial samples, one set of supports is held fixed, while the other set of supports is displaced in the orthogonal direction. This generates a hole in the Elastosil with an elliptical shape (Fig.~\ref{figS1}(c)). For the case of isotropic tension, the strain is found by $\epsilon = (d_\mathrm{f} - d_\mathrm{i})/d_\mathrm{i}$, where $d_\mathrm{i}$ and $d_\mathrm{f}$ are the initial (i.e. the state after the pre-strain has been relieved) and final diameters of the free-standing Elastollan film. In the case of biaxial tension, the strain along the high-tension direction is evaluated as $\epsilon_\mathrm{high} = (d_\mathrm{high,f}-d_\mathrm{i})/d_\mathrm{i}$, where $d_\mathrm{high,f}$ is the final diameter of the free-standing Elastollan film along the high-tension axis.

\begin{figure}[]
\begin{center}
     \includegraphics[width=8.9cm]{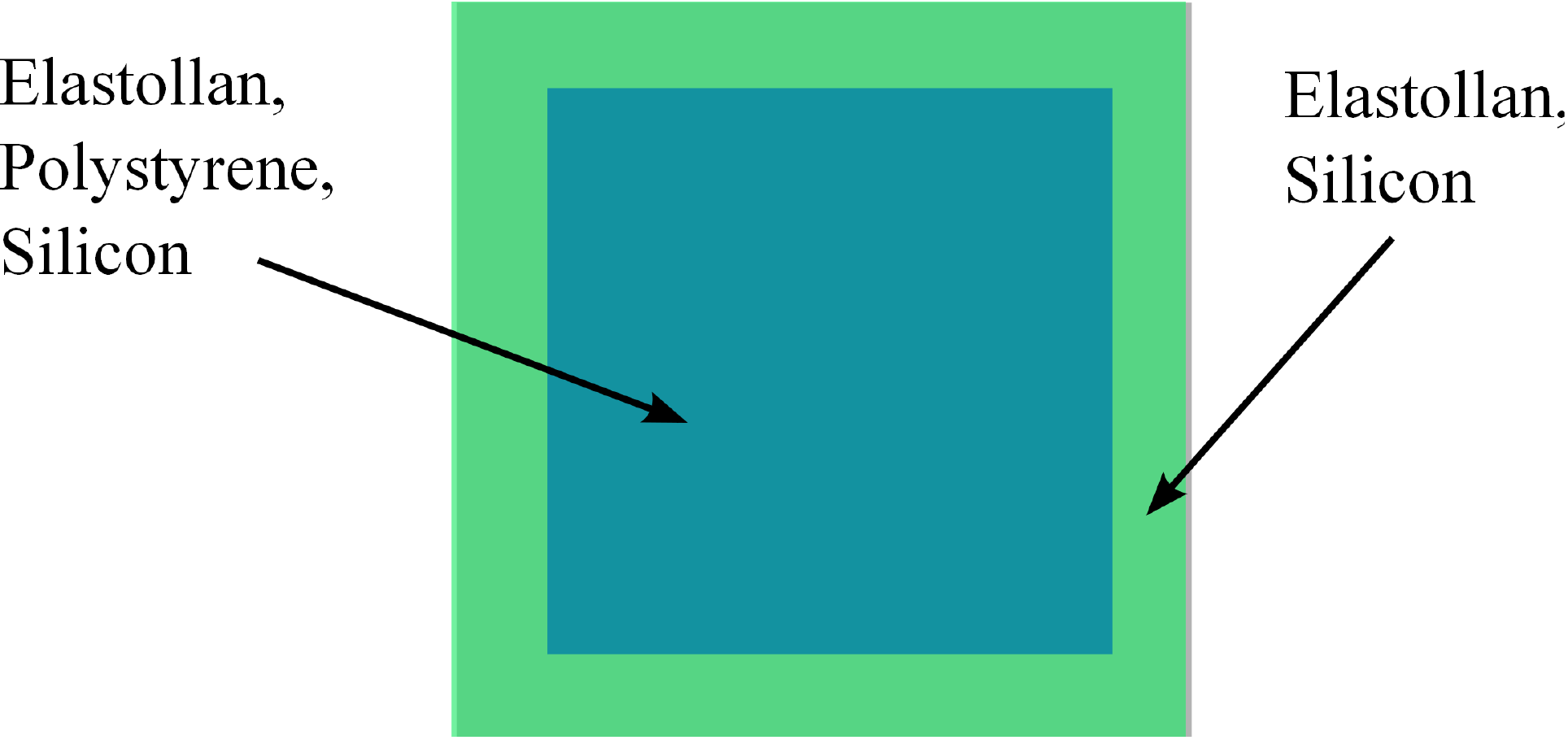}
\end{center}
\caption{Top view schematic of the final sample. The labels indicate the different material layers of the sample, listed in order from the top surface to the bottom surface. The ellipsometry measurements are performed in the region of Elastollan on silicon.}
\label{figS2}
\end{figure}

Next, the PS sample is brought into contact with the strained Elastollan film. The Elastollan adheres strongly to the PS and also to the bare silicon frame where the PS has been removed. Using a scalpel, the excess Elastollan is cut to free the sample from the straining set up. At this point, the sample looks as depicted in Fig.~\ref{figS2}. Using ellipsometry, the final Elastollan film thickness $h$ is measured in the portion of the sample with no PS. The sample is then ready for the experiment.

\section{Mechanical Relationships for Straining}
Using Hooke's law~\cite{timoshenko1951}, we can predict simple stress-strain relationships for our films under the assumption of Hookean elasticity. We assume that there is no stress acting in the direction orthogonal to the film ($z$-direction), i.e. $\sigma_z = 0$. We also know that the mechnical tension is related to stress through film thickness $T = h  \sigma$. As such, we may derive a simple  expression for isotropic tension generated upon straining:

\begin{equation}
T = \frac{Eh\epsilon}{1-\nu},
\label{T}
\end{equation}
where $\nu$ is the Poisson ratio of the elastomer, which can be assumed to be 0.5, and $E$ is the Young's modulus. Thus, $\epsilon h$ is an appropriate indicator for the tension in our films. For the case of biaxial tension where the film is held fixed along one direction ($\epsilon_\mathrm{low} = 0$), we arrive at:

\begin{equation}
T_\mathrm{low} = \frac{Eh \nu \epsilon_\mathrm{high}}{1-\nu^2},
\label{T_low}
\end{equation}
\begin{equation}
T_\mathrm{high} = \frac{Eh\epsilon_\mathrm{high}}{1-\nu^2},
\label{T_high}
\end{equation}
where ``high" and ``low" indicate the high and low tension directions. Here, we see that under the assumption of Hookean elasticity, $T_\mathrm{high} = 2T_\mathrm{low}$, since $\nu = 0.5$. In the biaxial tension experiments, $\epsilon_\mathrm{high} \sim$~100\%, $h\sim$ 100~nm, and $E\sim 10^7$~Pa~\cite{elastollan_modulus3}, which leads to $T_\mathrm{high} \sim$ 1.3~N/m. Thus, it is clear that mechanical tension will dominate over interfacial tensions in determining the Laplace pressure in the rim.

Since volume is conserved in a material with $\nu = 0.5$, it is possible to predict the final thickness upon an isotropic strain:
\begin{equation}
h = \frac{h_0}{(1+\epsilon)^2},
\label{h}
\end{equation}
as well as for a biaxial strain with $\epsilon_\mathrm{low} = 0$:
\begin{equation}
h = \frac{h_0}{(1+\epsilon_\mathrm{high})}.
\label{h_bi}
\end{equation}
Using the equations above, the initial film thickness $h_0$ was chosen to produce the desired $h$ (50 nm or 100 nm) once strained.

\section{Equilibrium contact angle}
\begin{figure}[]
\begin{center}
     \includegraphics[width=8.9cm]{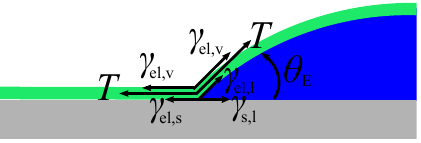}
\end{center}
\caption{Balance of interfacial and mechanical tensions to calculate the equilibrium contact angle that the rim subtends with the substrate.}
\label{figS3}
\end{figure}

To calculate the equilibrium contact angle that the rim subtends with the substrate, we appeal to a balance of the interfacial and mechanical tensions depicted in Fig.~\ref{figS3}. In this picture, we employ the simplifying assumption that the formation of holes does not alter the mechanical tension in the film. Carrying out the force balance in the horizontal direction yields:

\begin{equation}
T+\gamma_\mathrm{el,v}+\gamma_\mathrm{el,s} = \big( T + \gamma_\mathrm{el,v}+\gamma_\mathrm{el,l}\big) \mathrm{cos}\theta_\mathrm{E} +\gamma_\mathrm{s,l}.
\end{equation}
From the equation above, using the small angle approximation in which $\mathrm{cos}\theta \approx 1- \theta^2/2$, it is straightforward to show that:

\begin{equation}
\theta_\mathrm{E} = \frac{\theta_0}{\sqrt{1+\frac{T}{\gamma_\mathrm{el,v}+\gamma_\mathrm{el,l}}}}.
\end{equation}

\section{Elongated holes at late times}
\begin{figure}[]
\begin{center}
     \includegraphics[width=8.9cm]{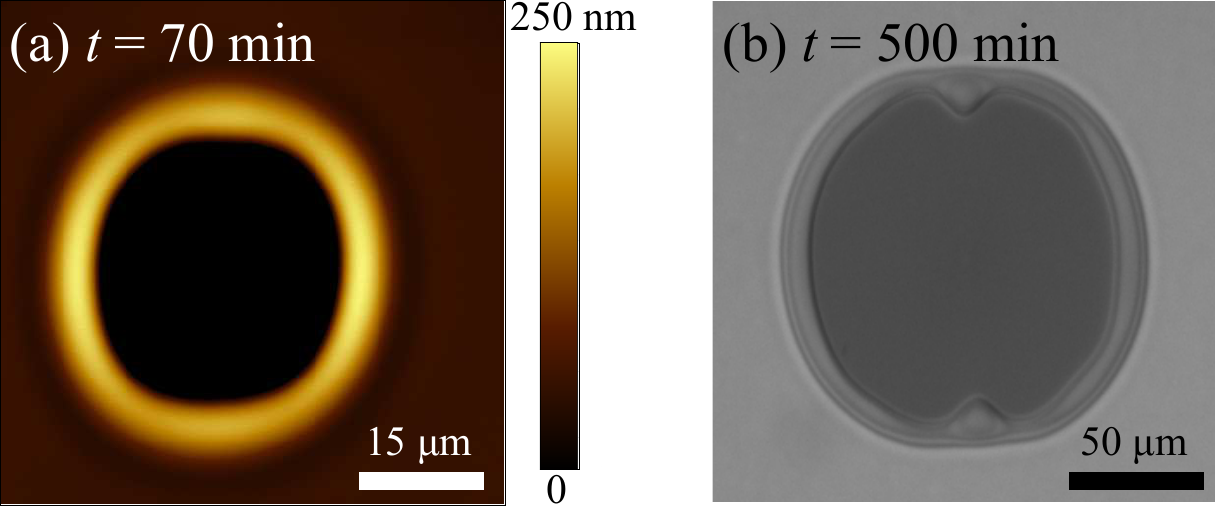}
\end{center}
\caption{(a) AFM scan of the hole from Fig.~3(b) once evidence of the rim instability sets in on the high tension sides of the hole. (b) Optical micrograph of the hole after the rim instability has started forming fingers at the high tension ends.}
\label{figS4}
\end{figure}

As the elongated holes continue to grow beyond the early stages described in the main manuscript, we observe that the rim instability rapidly begins to set in on the high tension side of the rim. The early stages of this instability can be seen in an AFM scan of the hole in Fig.~3(b) at $t =$ 70 min shown in Fig.~\ref{figS4}(a) where there is a bulge forming in the high tension side of the rim. The later stage of this instability is showcased by the optical image in Fig.~\ref{figS4}(b), where there are fingers forming at the high tension ends of the hole, yet the low tension side of the rim appears completely stable. In fact, the rim instability sets in at the high tension ends long before the stage of constant dewetting velocity has been reached. For this reason, the physics contained in Fig.~2(a), wherein holes exhibit a smaller $v$ when the tension is larger, cannot be applied to these anisotropic experiments. We suspect that the root of this effect is that depending on a liquid cylinder's orientation, the Plateau-Rayleigh instability can become enhanced or suppressed when a tension anisotropy is introduced in the elastomeric film capping the liquid -- this is beyond the scope of this study and will be investigated in future work.

\section{Supplemental Movies}
\noindent \textbf{Movie S1 - } Hole growth when the capping elastomer has an isotropic tension ($\epsilon h =$~8~nm). The movie is 750 min long.  \\
\textbf{Movie S2 - } Growth of square holes.

\end{document}